\documentclass{ws-ijmpe} 
\usepackage[super,compress]{cite}
\usepackage{amsmath}
\usepackage{amssymb}
\usepackage{graphicx}

\usepackage{ulem}
\usepackage{mdframed}
\usepackage{xcolor}  

\newcommand\qexpect[1]{ \langle #1 \rangle_q^{(\beta)}} 
\newcommand\qoneexpectphys[1]{ \langle #1 \rangle_{q=1}^{(\beta_{\mathrm{ph}})}} 
\newcommand\qoneexpectfreephys[1]{ \langle #1 \rangle_{q=1}^{(\beta_{\mathrm{ph}}), f}} 
\newcommand\qexpectfree[1]{ \langle #1 \rangle_q^{(\beta), f}} 
\newcommand\qexpectfreemassless[1]{ \left. \langle #1 \rangle_q^{(\beta), f} \right|_{\mathrm{massless}}} 
\newcommand{\Ueff}{U_{\mathrm{eff}}}
\newcommand{\betaph}{\beta_{\mathrm{ph}}}  
\newcommand{\Tph}{T_{\mathrm{ph}}}  
\newcommand{\Tdist}{T_{\mathrm{dist}}}  
\newcommand{\Tcrit}{T_{\mathrm{ph}, q}^{\mathrm{cr}}}  
\newcommand{\Tcritone}{T_{\mathrm{ph}, q=1}^{\mathrm{cr}}}  
\newcommand{\Tr}{\mathrm{Tr}}
\title{Chiral phase transition within the linear sigma model in the Tsallis nonextensive statistics based on density operator}
\author{Masamichi Ishihara}
\address{Department of Human Life Studies,\\
 Koriyama Women's University,\\
Koriyama, Fukushima, 963-8503, Japan\\
m\_isihar@koriyama-kgc.ac.jp
}

\begin{document}
\maketitle
\begin{abstract}
We studied the chiral phase transition for small $|1-q|$ within the Tsallis nonextensive statistics of the entropic parameter $q$,  
where the quantity $|1-q|$ is the measure of the deviation from the Boltzmann-Gibbs statistics. 
We adopted the normalized $q$-expectation value in this study. 
We applied the free particle approximation and the massless approximation in the calculations of the expectation values. 
We estimated the critical physical temperature, 
and obtained the chiral condensate, the sigma mass, and the pion mass,
as functions of the physical temperature $T_{\mathrm{ph}}$ for various $q$.  
We found the following facts.
The $q$-dependence of the critical physical temperature is $1/\sqrt{q}$.  
The chiral condensate at $q$ is smaller than that at $q'$ for $q>q'$.
The $q$-dependence of the pion mass and that of the sigma mass reflect the $q$-dependence of the condensate.
The pion mass at $q$ is heavier than that at $q'$ for $q>q'$.
The sigma mass at $q$ is heavier than that at $q'$ for $q>q'$ at high physical temperature, 
while the sigma mass at $q$ is lighter than that at $q'$ for $q>q'$ at low physical temperature. 
The quantities which are functions of the physical temperature $\Tph$ and the entropic parameter $q$
are described by only the effective physical temperature defined as $\sqrt{q} T_{\mathrm{ph}}$ under the approximations. 
\ccode{25.75.Nq, 11.30.Rd, 25.75.-q, 05.70.Fh}
\end{abstract}

\section{Introduction}
Power-like distributions appear in various branches of science.
The Tsallis nonextensive statistics \cite{Book:Tsallis} is a statistical mechanics which deals with power-like distributions.
The Tsallis nonextensive statistics is one parameter extension of the Boltzmann-Gibbs statistics,  
and the statistics has two parameters: the temperature $T$ and the entropic parameter $q$.
The deviation from the Boltzmann-Gibbs statistics is measured by the quantity $|1-q|$. 
The statistics has been applied to various phenomena which show the power-like distributions.

Momentum distributions in high energy collisions have been studied by many researchers. 
It was shown that the momentum distributions at high energies are described well by Tsallis-type distributions 
\cite{Alberico2009, Cleymans2012, Marques2015, GS2015, Azmi2015, Zheng2016, Thakur-AHEP2016, Lao2017, Cleymans2017-WoC, Cleymans2017, Yin2017, Osada-Ishihara-2018, Bhattacharyya2017-prepri,  Si-AHEP2018} .
In these studies, the values of $q$ were estimated, and the values of $|1-q|$ are close to $0.1$.
The deviation from the Boltzmann-Gibbs distribution is small at high energies, 
and the effects of the nonextensivity on physical quantities, such as correlation and fluctuation, for small $|1-q|$ have been studied
\cite{Alberico2000, Ishihara2017-corr-diffmass, Ishihara2017-fluctuation, Osada-Ishihara-2018}. 

Some origins of the Tsallis-type distribution are presented,  and one of them is the nonextensivity of the Tsallis entropy.
It is also considered that the fluctuation of the temperature and noise are the origins:
the temperature fluctuation cause the Tsallis distribution 
which was derived within the superstatistics: \cite{Beck-Cohen-2003, Beck-Continuum-2004}.
The cluster decay model in the Boltzmann-Gibbs statistics is also the origin of the Tsallis-type distribution \cite{Bialas-PLB-2015}.
The distribution was also derived in the special case of axiomatic statistics \cite{Hanel-EPL-2011}, 
and was used to study the momentum distributions and the particle ratios \cite{Tawfik-EPJA-2016, Tawfik-Chinese-2017, Tawfik-Indian-2018}. 
Many origins of the Tsallis-type distribution exist.

An interesting topic in high energy heavy ion collisions  is the chiral phase transition. 
The order parameter of the chiral symmetry is affected by the distribution.
It is required to study the chiral phase transition in a high energy collision in the presence of a Tsallis distribution,
because the momentum distribution is well described by the Tsallis distribution.
The Nambu-Jona-Lasinio (NJL) model \cite{Rozynek2009, Rozynek2016} and 
the linear sigma model \cite{Ishihara2015, Ishihara2016, Shen2017}
were used to study the chiral phase transition under the power-like distribution.
To study the effects of the Tsallis-type distribution within the Tsallis nonextensive statistics is meaningful,  
because the distribution of the particles related to the chiral symmetry is described by a power-like distribution. 
The study of the chiral phase transition in the presence of the power-like distribution is a significant topic at high energies.

It is not trivial that the (temperature-like) parameter introduced in the Tsallis nonextensive statistics corresponds to the temperature observed,  
and the ``physical'' temperature is often defined. 
The temperature $\Tdist$ in the Tsallis distribution is often called $q$-temperature \cite{Tawfik-Indian-2018} 
and the temperature defined by the relation between the entropy and the internal energy is called physical temperature $\Tph$
\cite{Abe-PLA2001,Suyari-PTPsupple2006}. 
It was shown that, for free particles, the distribution in the Tsallis nonextensive statistics with $\Tph$ 
is the same as the Tsallis distribution with $\Tdist$ for small $|1-q|$ 
when $\Tph$ is equal to $\Tdist$.
The meaning of the parameter introduced in a theory is not always trivial
even when the well-known principle (maximum entropy principle) is applied.

It is not easy to calculate quantities within the Tsallis nonextensive statistics, because of the power-like distribution.
In the statistics, quantities were calculated analytically \cite{Bhattacharyya2016} and estimated approximately. 
One way to calculate quantities is to expand the density operator as a series of the measure of the deviation $(1-q)$. 
For instance, the propagator was calculated with the $(1-q)$ expansion in quantum field theory \cite{Kohyama-Tsallis06}. 
The calculation with the density operator in a nonextensive system is required in the quantum statistics.

In this study,  we calculate the physical quantities with the density operator 
within the framework of the Tsallis nonextensive statistics for small $|1-q|$, assuming the realization of the Tsallis nonextensive statistics.
To study the realization of the Tsallis nonextensive statistics is out of scope in this paper.
We adopt the normalized $q$-expectation value \cite{Tsallis1998, Aragao-PhysicaA2003, Eicke-prepri, Kalyana2000}: 
the normalized $q$-expectation value of $1$ is equal to one. 
We study the chiral symmetry restoration in the Tsallis nonextensive statistics by employing the linear sigma model, 
because the fields are affected by power-like distributions. 
The physical temperature \cite{Book:Tsallis, Suyari-PTPsupple2006, Aragao-PhysicaA2003, Eicke-prepri, Kalyana2000, Abe-PLA2001, Toral-PhysicaA2003} 
dependence of a quantity is estimated for various $q$.
The results should be meaningful when the physical temperature is (almost) identical to the temperature observed with several functions
\cite{Adamczyk-PRC-2017}. 

This paper is organized as follows.
In Sec.~\ref{sec:tsallis-stat}, the Tsallis nonextensive statistics is briefly reviewed.
The normalized $q$-expectation value described with the physical temperature for small $|1-q|$, 
which was shown in the previous paper \cite{Ishihara-arXiv:1804.09144}, is presented.
In Sec.~\ref{sec:application}, the method in Sec.~\ref{sec:tsallis-stat} is applied to 
the linear sigma model to study the chiral phase transition.
The expression of the critical physical temperature is derived. 
In Sec.~\ref{sec:results}, 
the chiral condensate, sigma mass, and pion mass, are estimated as functions of the physical temperature for various $q$.
The last section is assigned for conclusion and discussion.

\section{The Tsallis nonextensive statistics for small $|1-q|$}
\label{sec:tsallis-stat}
\subsection{Brief review of the Tsallis nonextensive statistics}

The density operator in the Tsallis nonextensive statistics of the entropic parameter $q$ \cite{Book:Tsallis, Aragao-PhysicaA2003}
is given by 
\begin{align}
\rho &= \frac{1}{Z_q} \rho_u ,  \qquad 
\rho_u \equiv \left[ 1 - (1-q) \left( \frac{\beta}{c_q^{(\beta)}} \right) (H-\qexpect{H}) \right]^{1/(1-q)} ,  
\end{align}
where $Z_q$ is the partition function which is defined by $Z_q = \Tr \rho_u$, 
$\beta$ is the inverse temperature, 
$c_q^{(\beta)}$ is a factor,  $H$ is the Hamiltonian, 
and $\qexpect{H}$ is the normalized $q$-expectation value of the Hamiltonian.
We note that the quantity $c_q^{(\beta)}$ is given by $c_q^{(\beta)} = \Tr \rho^q$, 
and that the function $[1 - (1-q) x ]^{1/(1-q)}$ approaches $\exp(-x)$ as $q$ approaches one. 
The normalized $q$-expectation value of a quantity $A$ is defined by 
\begin{align}
\qexpect{A} := \Tr (\rho^q A)  / \Tr (\rho^q).
\end{align}
The normalized $q$-expectation value satisfies $\qexpect{1} = 1$.

The partition function $Z_q$ and the factor $c_q$ are related each other:
\begin{equation}
c_q^{(\beta)} = (Z_q)^{1-q} .
\label{cqZqrelation}
\end{equation}
This relation is easily derived by using the subsequent equations for the density operator: 
$\rho = (Z_{q-1})^{q-1} \rho^q [ 1 - (1-q) (\beta/c_q) (H-\qexpect{H}) ]$ and $\Tr \rho =1$.
The inverse physical temperature $\betaph$ is defined 
by using the inverse temperature $\beta$ and the quantity $c_q^{(\beta)}$ as follows: 
\begin{equation}
\betaph := \frac{\beta}{c_q^{(\beta)}} .
\label{inv-invphys-relation}
\end{equation}
These relations are basic in this study.
In the Tsallis nonextensive statistics, 
we should pay attention to the definition of the expectation value, besides the normalization.

\subsection{The calculation of the normalized $q$-expectation value for small $|1-q|$}
We focus on a quantity for small $|1-q|$.   
As shown in Ref.~\refcite{Ishihara-arXiv:1804.09144}, 
we attempt to calculate the quantity by expanding the normalized $q$-expectation value with respect to $\varepsilon \:= 1-q$.
The quantities $\qexpect{H}$, $c_q^{(\beta)}$, and $Z_q$ are expanded with respect to $\varepsilon$:
\begin{subequations}
\begin{align}
&\qexpect{H} = E_0 - \varepsilon E_1 + O(\varepsilon^2),
\\
&c_q^{(\beta)} = c_0^{(\beta)} - \varepsilon c_1^{(\beta)} + O(\varepsilon^2),
\\
&Z_q = Z_0 - \varepsilon Z_1 + O(\varepsilon^2).
\end{align}
\end{subequations}
It is shown from Eq.~\eqref{cqZqrelation} that the value of $c_0^{(\beta)}$ is one. 
The inverse temperature is rewritten by the inverse physical temperature from Eq.~\eqref{inv-invphys-relation}.
\begin{align}
\beta = \betaph - \varepsilon c_1^{(\betaph)} \betaph + O(\varepsilon^2) . 
\end{align}
The quantity $(\rho_u)^q$ is expanded as follows:
\begin{align}
(\rho_u)^q = e^{\beta E_0} e^{-\beta H} \left\{ 1+ \varepsilon \left[ L_0 + L_1 H + L_2 H^2 \right] + O(\varepsilon^2) \right\} ,
\end{align}
where $L_0$, $L_1$, and $L_2$ are given by 
\begin{subequations}
\begin{align}
L_0 &= - \beta \left[ (1-c_1^{(\beta)}) E_0 + E_1 + \frac{1}{2} \beta (E_0)^2 \right] ,\\
L_1 &= \beta \left[ (1-c_1^{(\beta)}) + \beta E_0 \right] ,\\
L_2 &= -\frac{1}{2} \beta^2  .
\end{align}
\end{subequations}

With these expressions, 
we obtain the normalized $q$-expectation value of the quantity $A$ to the $O(\varepsilon)$ 
with the inverse physical temperature \cite{Ishihara-arXiv:1804.09144}:
\begin{align}
\qexpect{A} = & \qoneexpectphys{A}
+ \varepsilon \Big\{ \betaph\left( 1 + \betaph E_0^{(\betaph)} \right) 
    \left[\qoneexpectphys{HA} - \qoneexpectphys{H} \qoneexpectphys{A} \right] 
\nonumber \\ & 
- \frac{1}{2} \left( \betaph \right)^2 \left[\qoneexpectphys{H^2A} - \qoneexpectphys{H^2} \qoneexpectphys{A} \right] \Big\} 
+ O(\varepsilon^2) . 
\label{final-eq-O-betaphys}
\end{align}
We use Eq.~\eqref{final-eq-O-betaphys} to calculate the normalized $q$-expectation values of fields.

\section{Application to the linear sigma model}
\label{sec:application}
\subsection{The normalized $q$-expectation value of the quadratic term of a scalar field}
We begin with the calculation of the normalized $q$-expectation value of the square of a scalar field $: \varphi_j^2 :$ with a free Hamiltonian $H^f$, 
where the notation $:A:$ represents the normal ordered quantity of $A$ and the index $j$ distinguishes the field.
The free Hamiltonian $H^f$ is 
\begin{align}
H^f &= \sum_{i=0}^{N-1} H_i^f , \qquad 
H_i^f := \int d\vec{x} \left[ \frac{1}{2} (\partial^0 \varphi_i)^2 +  \frac{1}{2} (\nabla \varphi_i)^2  + \frac{1}{2} m_i^2 (\varphi_i)^2 \right] . 
\end{align}
The normalized $q$-expectation value of the quantity $A$ under the free particle approximation, $\qexpectfree{A}$,
is calculated with the free Hamiltonian $H^f$. 
The field $\varphi_j$ is expanded as follows:
\begin{align}
\varphi_j = \sum_{\vec{k}} \left. \frac{1}{\sqrt{2\omega_{j\vec{k}}V}}   \left( a_{j\vec{k}} e^{-ikx} + a_{j\vec{k}}^{\dag} e^{ikx} \right) \right|_{k^0 =\omega_{j \vec{k}}}
,
\end{align}
where $\omega_{j\vec{k}}$ is the energy of a free particle, 
and $a_{j\vec{k}}$ is the annihilation operator with the commutation relations: 
$[a_{i\vec{k}},  a_{j\vec{l}}^{\dag} ] = \delta_{ij}\delta_{\vec{k},\vec{l}}$ and $[a_{i\vec{k}},  a_{j\vec{l}} ] = 0$.

The normalized $q$-expectation value of $a_{i\vec{k}} a_{j\vec{l}}$ under the free particle approximation, $\qexpectfree{a_{i\vec{k}} a_{j\vec{l}}}$, is zero.
The quantity $\qexpectfree{a^{\dag}_{i\vec{k}} a_{j\vec{l}}}$ is also zero for $i \neq j$.
We focus on the calculation of $\qexpectfree{a^{\dag}_{j\vec{k}} a_{j\vec{l}}}$ at $\vec{k} = \vec{l}$ with Eq.~\eqref{final-eq-O-betaphys}, 
because the expectation value is zero for $\vec{k} \neq \vec{l}$.


The free Hamiltonian is divided into two parts to calculate $\qexpectfree{a^{\dag}_{j\vec{k}} a_{j\vec{k}}}$ as follows:
\begin{align}
H^f = \omega_{s\vec{k}} n_{s\vec{k}}  + H^{R} , 
\end{align}
where $n_{s\vec{k}} := a^{\dag}_{s\vec{k}} a_{s\vec{k}}$ and  $H^R$ is the remaining part. 
We obtain
\begin{align}
\qexpectfree{n_{s\vec{k}}} 
= &
\qoneexpectfreephys{n_{s\vec{k}}} 
+ \varepsilon \Big\{ \betaph\left( 1 + ( \betaph \omega_{s\vec{k}} ) \qoneexpectfreephys{n_{s\vec{k}}} \right) 
\nonumber \\ & \quad \times 
    \left[\qoneexpectfreephys{(n_{s\vec{k}})^2 } - \left( \qoneexpectfreephys{n_{s\vec{k}}} \right)^2 \right] 
\nonumber \\ &  \quad 
- \frac{1}{2} \left( \betaph \omega_{s\vec{k}} \right)^2 
\left[\qoneexpectfreephys{(n_{s\vec{k}})^3} - \qoneexpectfreephys{(n_{s\vec{k}})^2} \qoneexpectfreephys{n_{s\vec{k}}} \right] \Big\} 
\nonumber \\ & \qquad 
+ O(\varepsilon^2) . 
\label{eq:nsk-freeHamiltonian-betaphys}
\end{align}
Equation~\eqref{eq:nsk-freeHamiltonian-betaphys} gives 
\begin{align}
\qexpectfree{n_{s\vec{k}}} 
&= \frac{1}{\exp(\betaph \omega_{s\vec{k}}) -1} 
+ \varepsilon \Bigg\{ 
        \frac{ (\betaph \omega_{s\vec{k}})  \exp(\betaph \omega_{s\vec{k}}) }{\left[ \exp(\betaph \omega_{s\vec{k}}) -1 \right]^2} 
\nonumber \\ & \quad 
       - \frac{(\betaph \omega_{s\vec{k}})^2 \exp(\betaph \omega_{s\vec{k}}) \left[ \exp(\betaph \omega_{s\vec{k}}) +1 \right] }
             {2 \left[ \exp(\betaph \omega_{s\vec{k}}) -1 \right]^3} 
        \Bigg\}
+ O(\varepsilon^2) 
. 
\label{q-1expansion-of-nsk}
\end{align}
This form for multiple fields under the free particle approximation, Eq.~\eqref{q-1expansion-of-nsk},
coincides with the form given in Ref.~\refcite{Ishihara-arXiv:1804.09144} for one field.

We obtain the quantity $\qexpectfree{:(\varphi_j)^2:}$ to the $O(\varepsilon)$ with Eq.~\eqref{q-1expansion-of-nsk}:
\begin{align}
& \qexpectfree{:(\varphi_j)^2:} \nonumber \\ 
& = 
\frac{1}{V} \sum_{\vec{k}} \left( \frac{1}{\omega_{j\vec{k}}} \right) 
\left\{ 1 + \varepsilon \left[ (\betaph \omega_{j\vec{k}}) - \frac{1}{2} (\betaph \omega_{j\vec{k}})^2  \right] \right\}  
\frac{1}{\left[ \exp(\betaph \omega_{j\vec{k}}) -1 \right]} 
\nonumber \\ & \quad  + 
\frac{1}{V} \sum_{\vec{k}}  \left( \frac{1}{\omega_{j\vec{k}}} \right)
\varepsilon \left[ (\betaph \omega_{j\vec{k}}) - \frac{3}{2} (\betaph \omega_{j\vec{k}})^2  \right] 
\frac{1}{\left[ \exp(\betaph \omega_{j\vec{k}}) -1 \right]^2} 
\nonumber \\ & \quad  + 
\frac{1}{V} \sum_{\vec{k}} \left( \frac{1}{\omega_{j\vec{k}}} \right)
(-\varepsilon) (\betaph \omega_{j\vec{k}})^2  \frac{1}{\left[ \exp(\betaph \omega_{j\vec{k}}) -1 \right]^3} 
\nonumber \\ & \quad  + 
O(\varepsilon^2) . 
\label{eqn:freeHamiltoninan:varphi2}
\end{align}
The right-hand side of Eq.~\eqref{eqn:freeHamiltoninan:varphi2} is represented by integrals
when the large volume limit is taken. 
We apply the massless approximation that the masses are ignored in the integrals
\cite{Ishihara2017-corr-diffmass, Gavin1994, Ishihara1999, Ishihara-IJMPA33}.
After these procedures, we obtain the expression of \mbox{$\qexpectfree{:(\varphi_j)^2:}$} under the massless approximation:
\begin{align}
\qexpectfreemassless{:(\varphi_j)^2:}  
& = \frac{1}{2\pi^2 \betaph^2} 
\Bigg\{ F(1,0,1;0)  
+ \varepsilon \Big[ F(1, 0, 2; 0) - \frac{1}{2} F(1, 0, 3; 0) 
\nonumber \\ & \qquad 
+ F(2, 0, 2; 0) - \frac{3}{2} F(2, 0, 3; 0) - F(3, 0, 3; 0) \Big] \Bigg\}
\nonumber \\ & \qquad 
+ O(\varepsilon^2), 
\label{eqn:brief-representation}
\end{align}
where $F(a, b, c; \nu)$ is defined by 
\begin{align}
F(a, b, c; \nu) := \int_0^{\infty} dx x^c \frac{(e^x)^b}{(e^{x+\nu} -1)^a}  . 
\end{align}
The integrals that appear in Eq.~\eqref{eqn:brief-representation} \cite{Gradshteyn, Ishihara-IJMPA33} are given explicitly:
\begin{subequations}
\begin{align}
F(1, 0, c; 0) &= \Gamma(c+1) \zeta(c+1), \\
F(2, 0, c; 0) &= \Gamma(c+1) \left[ \zeta(c) - \zeta(c+1) \right],\\
F(3, 0, c; 0) &= \frac{1}{2} \Gamma(c+1) \left[ \zeta(c-1) - 3 \zeta(c) + 2 \zeta(c+1) \right] ,
\end{align}
\label{eqn:results-integrals}
\end{subequations}
where $\Gamma(x)$ is the gamma function and $\zeta(x)$ is the zeta function. 
We finally obtain the expression of $\qexpectfree{:(\varphi_j)^2:}$ under the massless approximation
by using Eq.~\eqref{eqn:results-integrals}: 
\begin{align}
\qexpectfreemassless{:(\varphi_j)^2:}
&= \frac{1}{12 (\betaph)^2} \left[ 1 - \varepsilon + O(\varepsilon^2) \right] 
= \frac{q}{12 (\betaph)^2}  + O(\varepsilon^2) 
. 
\label{eqn:free:massless:varphi2}
\end{align}
We note that the most right-hand side of Eq.~\eqref{eqn:free:massless:varphi2} does not depend on the index $j$.
We use the above result, Eq.~\eqref{eqn:free:massless:varphi2}, in the next subsection.

\subsection{The condensate, the effective particle mass, and the critical physical temperature} 
The Hamiltonian density of the linear sigma model is 
\begin{equation}
{\cal H} = \frac{1}{2} (\partial^0 \phi)^2  + \frac{1}{2} (\nabla \phi)^2  + \frac{\lambda}{4} (\phi^2 - v^2)^2 - G \phi_0, 
\end{equation}
where $\phi \equiv (\phi_0, \phi_1,  \cdots, \phi_{N-1})$, 
$\phi^2 \equiv \sum_{j=0}^N \phi_j^2$, and  $(\partial^i \phi)^2 \equiv \sum_{j=0}^N (\partial^i \phi_j)^2$.
The quantities, $\lambda$, $v$, and $G$, are the parameters of the linear sigma model.
We shift the fields as $\phi_j = \phi_{j\mathrm{c}} + \varphi_j$,
where the quantity $\phi_{j\mathrm{c}}$ is the condensate of the field $\phi_j$. 
The normal ordered Hamiltonian density is given by 
\begin{align}
:{\cal H}: &= 
\frac{\lambda}{4} \left( \phi_c^2 - v^2 \right)^2 - G \phi_{0c} 
+ : \frac{1}{2} (\partial^0 \varphi)^2  + \frac{1}{2} (\nabla \varphi)^2  : 
+ \left[ \lambda (\phi_c^2 - v^2)  ( \phi_c \cdot \varphi) - G \varphi_0 \right] 
\nonumber \\ & \qquad 
+ \left[ \frac{\lambda}{2} (\phi_c^2 - v^2) : ( \varphi^2):   + \lambda : (\phi_c \cdot \varphi) ^2: \right] 
+ \lambda :(\phi_c \cdot \varphi)  (\varphi^2) : 
+ \frac{\lambda}{4} \left[ : (\varphi^2)^2 : \right]
,
\end{align}
where it is assumed that $\phi_{j \mathrm{c}}$ is uniform and independent of  time. 
The representation $(\phi_c \cdot \varphi)$ indicates $\sum_{j=0}^{N-1} \phi_{j\mathrm{c}} \varphi_j$.

We calculate the normalized $q$-expectation value of ${\cal (:H:)}$. 
The $q$-expectation value under the free particle approximation is 
\begin{align}
\qexpectfree{:{\cal H}:} 
&= 
\frac{\lambda}{4} \left( \phi_c^2 - v^2 \right)^2 - G \phi_{0c} 
+ \qexpectfree{ : \frac{1}{2} (\partial^0 \varphi)^2 + \frac{1}{2} (\nabla \varphi)^2 : }
\nonumber \\ & \quad 
+ \frac{\lambda}{2} (\phi_c^2 - v^2) \qexpectfree{: ( \varphi^2):}  + \lambda \sum_{j=0}^{N-1}(\phi_{jc})^2  \qexpectfree{: (\varphi_j) ^2:} 
\nonumber \\ & \quad 
+ \frac{\lambda}{4} \qexpectfree{  : (\varphi^2)^2 : }
.
\label{eqn:H:qexpectfree}
\end{align}

We apply the massless approximation to the normalized $q$-expectation values in Eq.~\eqref{eqn:H:qexpectfree}, 
and obtain the effective potential $\Ueff$:  
\begin{align} 
\Ueff &\equiv \frac{1}{V} \int d\vec{x} \qexpectfreemassless{:{\cal H}:}
\nonumber \\ &
= \frac{\lambda}{4} \left( \phi_c^2 - v^2 \right)^2 - G \phi_{0c} 
+ \frac{\lambda}{2} (N+2) (\phi_{\mathrm{c}}^2)  I_q^{(\beta)}  + \Ueff^R
,
\end{align} 
where  $\Ueff^R$ represents the terms which are independent of the condensates.
The quantity $I_q^{(\beta)}$ is defined by 
\begin{align}
I_q^{(\beta)}  := \qexpectfreemassless{:(\varphi_j)^2:} .
\end{align}
The quantity $I_q^{(\beta)}$ is given explicitly in Eq.~\eqref{eqn:free:massless:varphi2}.

The condensate $\phi_{0\mathrm{c}}$ at the physical temperature $\Tph$ satisfies 
the equation $\partial \Ueff/\partial \phi_{0\mathrm{c}} = 0$ with $\phi_{j\mathrm{c}}=0$  $(j \neq 0)$. 
That is
\begin{align}
\lambda \left[ (\phi_{0c})^2  + (N+2) I_q^{(\beta)} - v^2 \right] \phi_{0\mathrm{c}} - G = 0 
.
\label{eqn:cond:phi0c}
\end{align}
We denote the solution of Eq.~\eqref{eqn:cond:phi0c} as $\overline{\phi}_{0\mathrm{c}}$. 
The effective mass squared $(M_j)^2$ on the vacuum which satisfies Eq.~\eqref{eqn:cond:phi0c} is given by
\begin{align}
(M_j)^2 &= \lambda \left[ (1 + 2 \delta_{j0} ) (\overline{\phi}_{0\mathrm{c}}) + (N+2) I_q^{(\beta)} -v^2 \right]
.
\end{align}

The critical physical temperature $\Tcrit$ is defined as the physical temperature 
at which a local minimum and a local maximum marge. 
The critical physical temperature is given by
\begin{align}
& \Tcrit = \left( \frac{1}{\sqrt{q}} \right) \Tcritone,  \label{eq:Tcrit:qdep} \\ 
& \Tcritone = \sqrt{\frac{12}{(N+2)} \left[ v^2 - \frac{3}{4} \left( \frac{4G}{\lambda} \right)^{2/3} \right]}  .  \nonumber 
\label{eq:Tcrit:qdep}
\end{align}
The critical physical temperature decreases as a function of $q$ according to Eq.~\eqref{eq:Tcrit:qdep}.

\section{Results}
\label{sec:results}
We calculated the physical temperature dependences of the quantities for various $q$:  
the physical temperature dependence of the condensate, that of the sigma mass, and that of the pion mass.
The values of the parameters were set as follows.
The number of the field $N$ was set to four.
The field for $j=0$ is the sigma field, and the fields for $j \neq 0$ are the pion fields.
The parameters of the linear sigma model, $\lambda$, $v$, and $G^{1/3}$, were set to 20, 87.4 MeV, and 119 MeV, respectively.  
These values generate $M_0=600$ MeV and $M_j=135$ MeV $(j \neq 0)$ at zero physical temperature. 
We studied the chiral phase transition with these values.

\begin{figure}
\begin{center}
\includegraphics[width=0.5\textwidth]{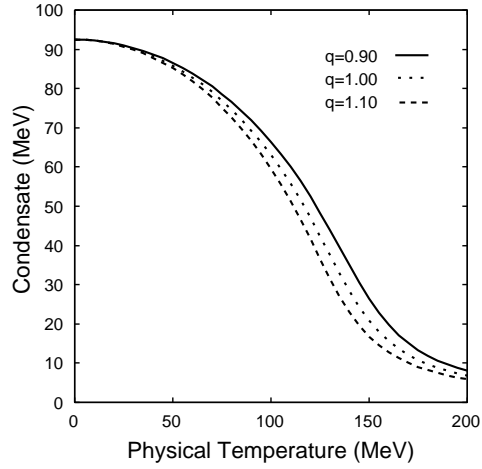}
\end{center}
\caption{Physical temperature dependence of the condensate for $q=0.9, 1.0$, and $1.1$. }
\label{Fig:condensate}
\end{figure}

Figure~\ref{Fig:condensate} shows the condensate as a function of the physical temperature $\Tph$  for $q=0.9, 1.0$, and $1.1$.
The condensate at $q$ is smaller than that at $q'$ for $q>q'$. 
This behavior is easily explained by the fact that the value of $I_q^{(\beta)}$  increases as $q$ increases.
The value of the condensate decreases faster with increasing $\Tph$ for large $q$.

We compare the present results in Fig.~\ref{Fig:condensate} with the results given in Ref.~\refcite{Shen2017}. 
In Ref.~\refcite{Shen2017}, the linear sigma model with quarks was employed and some quantities were calculated in the path integral method. 
They calculated the constituent (anti)quark mass as functions of the temperature at vanishing chemical potential  for some values of $q$. 
The constituent quark mass is a measure of the chiral symmetry restoration, and the mass is small when the chiral symmetry is restored. 
The $q$-dependence of the constituent quark mass in their study is similar to the $q$-dependence of the chiral condensate in the present study.
As for the chiral symmetry restoration, the present results are qualitatively consistent with those in Ref.~\refcite{Shen2017}.

\begin{figure}
\begin{center}
\includegraphics[width=0.5\textwidth]{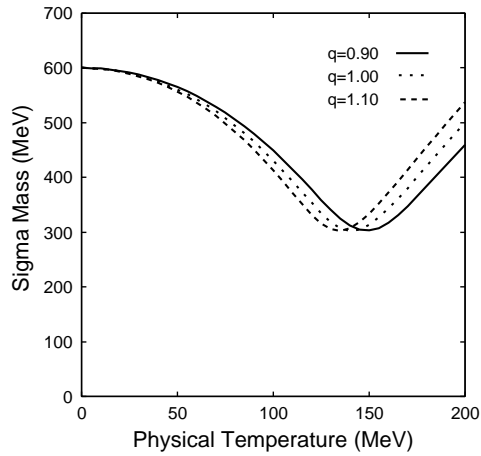}
\end{center}
\caption{Physical temperature dependence of the sigma mass for $q=0.9, 1.0$, and $1.1$. }
\label{Fig:sigma-mass}
\end{figure}

Figure~\ref{Fig:sigma-mass} shows the sigma mass as a function of $\Tph$ for $q=0.9, 1.0$, and $1.1$.
The sigma mass includes the effects of the nonextensivity through the condensate. 
Generally, the sigma mass decreases, reaches minimum, and increases after that, as the (physical) temperature increases.  
The $q$-dependence of the sigma mass shown in the figure is explained by the $q$-dependence of the condensate.
Therefore, the sigma mass at $q$ is lighter than that at $q'$ for $q>q'$ at low physical temperature.
Contrarily, the sigma mass at $q$ is heavier than that at $q'$ for $q>q'$ at high physical temperature.

\begin{figure}
\begin{center}
\includegraphics[width=0.5\textwidth]{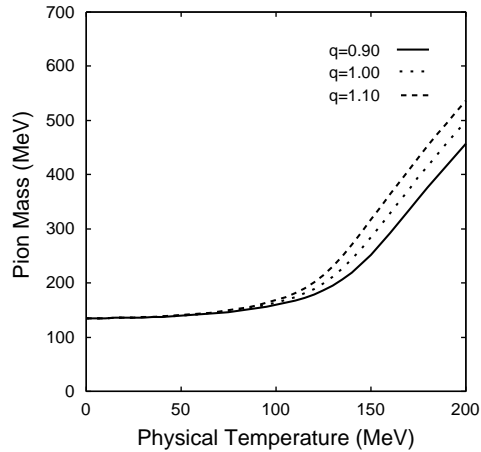}
\end{center}
\caption{Physical temperature dependence of the pion mass for $q=0.9, 1.0$, and $1.1$. }
\label{Fig:pion-mass}
\end{figure}

Figure~\ref{Fig:pion-mass} shows the pion mass as a function of $\Tph$  for $q=0.9, 1.0$, and $1.1$.
The pion mass also includes the effects of the nonextensivity through the condensate. 
The $q$-dependence of the pion mass is simple: the pion mass at $q$ is heavier than the mass at $q'$ for $q>q'$.

These $q$-dependences in the present results are reasonable, 
because the tail of the distribution becomes long as $q$ increases. 
That is, the value of $I_q^{(\beta)}$ increases as $q$ increases. 
This fact indicates the followings.
We deal with the two sets of the physical temperature and the entropic parameters, $(\Tph, q)$ and $(\Tph',q')$. 
The relation, $\Tph' = \sqrt{q/q'} \Tph$, is satisfied from Eq.~\eqref{eqn:free:massless:varphi2}  
when the value of $I_q^{(\beta)}$  at $(\Tph, q)$ and the value at $(\Tph',q')$ are the same.
This relation means  $\Tph' < \Tph$ for $q'>q$.  
This implies that the value of a quantity at $(\Tph', q') $ and the value at $(\Tph,q)$ are same.
Graphically, the curve at $q'$ is shifted to the lower physical temperature, compared with the curve at $q$, for $q'>q$.  
The behavior of the quantity as a function of $\Tph$ calculated numerically in this section is valid.

Finally, we comment on the critical physical temperature.
As shown in the previous section, 
the critical physical temperature decreases monotonically with the entropic parameter $q$. 
The behavior  is naively expected, because of the long tail of the distribution for large $q$. 

\section{Conclusion and Discussion}

We studied the chiral phase transition within the Tsallis nonextensive statistics for small $|1-q|$. 
The statistics has  two parameters: the temperature $T$ and the entropic parameter $q$. 
We adopted the normalized $q$-expectation value.
The inverse physical temperature is defined as $\betaph = {\beta}/{c_q^{(\beta)}}$, 
where the factor $c_q^{(\beta)}$ is included in the density operator.
The free particle approximation was applied in the calculation of the normalized $q$-expectation value, 
and the massless approximation was applied in the calculation of the momentum integrals. 
We obtained the $q$-dependence of the critical physical temperature. 
We calculated the chiral condensates, the sigma mass, and the pion mass, 
as functions of the physical temperature $\Tph$ for various $q$.

We found that the $q$-dependence of the critical physical temperature is $1/\sqrt{q}$. 
The value of the chiral condensate at $q$ is smaller than that at $q'$ for $q>q'$. 
This indicates that the chiral symmetry restoration occurs at low physical temperature for large $q$. 
The pion mass at $q$ is heavier than that at $q'$ for $q>q'$.
The sigma mass at $q$ is heavier than that at $q'$ for $q>q'$ at high physical temperature, 
while the sigma mass at $q$ is lighter than that at $q'$ for $q>q'$ at low physical temperature. 
These behaviors of the masses reflect the $q$-dependence of the chiral condensate.

The effects of the nonextensivity are included through the normalized $q$-expectation value of the square of the field \mbox{$\qexpect{:(\varphi_j)^2:}$}.
The quantities calculated in this study are described  without $q$ when the effective physical temperature $T_{\mathrm{ph}}^{\mathrm{eff}}$ is used, 
where $T_{\mathrm{ph}}^{\mathrm{eff}}$ is defined by $\sqrt{q} \Tph$,  
because $\qexpectfree{:(\varphi_j)^2:}$ under the massless approximation (see Eq.~\eqref{eqn:free:massless:varphi2}) 
is described as $(T_{\mathrm{ph}}^{\mathrm{eff}})^{2}/12 + O((1-q)^2)$.
The behavior of the quantity as a function of the physical temperature is easily explained by 
the behavior of the normalized $q$-expectation value of the square of the field. 
We note that the results obtained in this study agree with those obtained in the previous study of the $\phi^4$ theory \cite{Ishihara-IJMPA33}.

We focused on the effects of the deviation from the Boltzmann-Gibbs statistics in this study.  
The qualitative nature of the chiral phase transition was revealed in the Tsallis nonextensive statistics,
though some thermal effects were ignored.
The calculation will be improved by taking these effects into account.
For example, it will be possible to include some diagrams such as daisy diagrams 
when the effective masses in the momentum integrals are included. 
A gap equation for the effective mass will appear and some diagrams can be included by solving the gap equation. 
The calculation of the normalized $q$-expectation value of the square of the field, $\qexpectfree{:(\varphi_j)^2:}$, 
without the approximations used in this study 
may be required to calculate a quantity precisely as a function of the physical temperature for various $q$ in the future.

The author hopes that this work will be helpful for the readers to study the effects of the nonextensivity 
within the framework of the Tsallis nonextensive statistics.



\end{document}